\documentstyle[epsf]{l-aa}

\begin{document}

   \thesaurus{11         
              (11.09.1;
               11.09.2;
               11.11.1)}
   \title{NGC\,488: Has its massive bulge been build up by minor mergers?}

   \author{Burkhard Fuchs}


   \institute{Astronomisches Rechen-Institut Heidelberg,
              M\"onchhofstr. 12-14, 69120 Heidelberg, Germany}

   \date{Received; accepted  }

   \maketitle

   \begin{abstract}

   Using recent observations of the kinematics of the disk stars the
   dynamical state of the disk of NGC\,488 is discussed. The disk is
   shown to be dynamically `cool', so that NGC\,488 cannot have
   experienced many -- even minor -- mergers in the past.

   \end{abstract}

%

\section{Introduction}

It has been argued (Jore, Haynes \& Broeils 1997) that some of the massive
bulges of Sa type galaxies have not been formed during the collapse of the
protogalaxy or by secular evolution of the galactic disk (Pfenniger, Combes \&
Martinet 1994), but by capture of satellite
galaxies. Jore, Haynes \& Broeils (1997) present a sample of Sa galaxies with
kinematically distinct components in their inner parts such as counter-rotating
disks. These might be well interpreted as debris of satellite galaxies
which disintegrated while they merged with their parent galaxies. On the other
hand, Ostriker (1990) has pointed to the fact that the disks of Sa galaxies are
dynamically cool enough to develop spiral structure. Since galactic disks are
dynamically heated during the merging process, this sets severe constraints on
the accretion rate of satellites (T\'{o}th \& Ostriker 1992).

Recently Gerssen, Kuijken \& Merrifield (1997; hereafter referred to as GKM)
have observed the kinematics of the stellar disk of NGC\,488. NGC\,488 is a
typical Sa galaxy which is actually surrounded by dwarf satellites (Zaritsky et
al. 1993), so that there might have been indeed minor merger events in the
past. GKM discuss implications of their observations, in particular the shape
of the velocity ellipsoid, for dynamical disk heating by molecular clouds and
transient spiral density waves. But their data allow also to state Ostriker's
(1990)  objections to the scenario of steady accretion of small satellites in a
quantitative way. For this purpose I construct in sections 2 and 3 dynamical
models of the bulge and disk of NGC\,488 using the photometric and kinematical
data presented by GKM. In the final section I discuss the
dynamical state of the disk and implications for the merging history of
NGC\,488.
%
   \begin{figure}[htbp]
   \begin{center}
   \leavevmode
      \epsffile{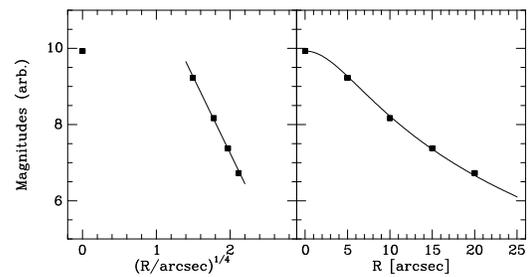}
      \caption{Surface brightness profile of the bulge of NGC\,488
      according to the bulge -- disk decomposition by GKM. The left panel
      shows the surface brightness profile as a de Vaucouleurs profile,
      while the right panel shows a surface brightness profile according
      to equation (1) fitted to the data ($\Box$).}
         \label{bulge}
   \end{center}
   \end{figure}
%
%

\section{Bulge -- disk decomposition}

I adopt the decomposition of the major-axis surface brightness profile into
bulge
and exponential disk contributions, respectively, given by GKM. In Fig. 1 the
surface brightness profile of the bulge is shown either as a de Vaucouleurs
profile (left panel) or the surface brightness profile of a -- spherical --
bulge model with a density distribution of the form
\begin{equation}
\rho_{\rm b} =
\rho_{\rm b0} \left( 1 + {\frac{r^2}{r_{\rm c,b}^2}}
\right)^{-\frac{3.5}{2}}\,.
\end{equation}
The surface brightness profile follows a similar law with the exponent lowered
by one and is shown in the right panel of Fig.~1 fitted to the data of GKM. The
core radius which I find this way is $r_{\rm c,b}$ = 6\farcs3 or
920 pc if a distance of 30 Mpc to NGC\,488 is assumed.

\section{Decomposition of the rotation curve}

GKM have determined the rotation curve of NGC\,488 using absorption line
spectra of the stars and give a parametrization in the form
\begin{equation}
v_c(R) = 336 \left(\frac{R}{R_{40}}\right)^{0.21} km/s\, ,
\end{equation}
where the fiducial radius $R_{40}$ corresponds to a galactocentric distance of
40\arcsec or 5.8 kpc. This is equal to the radial exponential scale length of
the disk in the B-band. The rotation curve determined by GKM fits well to the
rotation curve derived by Peterson (1980) using emission line spectra.
Peterson (1980) has also determined circular velocities in the region
dominated by the bulge, which I have combined to a single data point in Figs.
2 to 5. Unfortunately, no HI rotation curve is available.

To the observed rotation curve I fit a model rotation curve of the form
\begin{equation}
v_{\rm c}^2(R) = v_{\rm c,b}^2(R) + v_{\rm c,d}^2(R) + v_{\rm c,h}^2(R)\,,
\end{equation}
where $v_{\rm c,b}$, $v_{\rm c,d}$,
and $v_{\rm c,h}$ denote the contributions due to the
bulge, disk, and dark halo, respectively. The bulge contribution is given
according to equation (1) by
\begin{equation}
v_{\rm c,b}^2(R) = \frac{4\pi G\rho_{\rm b0}}{R} \int_0^R dr
r^2\left(1
+ {\frac{r^2}{r_{\rm c,b}^2}} \right)^{-\frac{3.5}{2}}
\end{equation}
where G denotes the constant of gravitation. \footnote{The integral can be
reduced to a hypergeometric function, although this is of little practical
advantage.}
The rotation curve of an infinitesimally thin exponential disk is given by
\begin{equation}
v_{\rm c,d}^2(R) = 4\pi G \Sigma_{\rm d0} h x^2\left(I_0(x)K_0(x) -
I_1(x)K_1(x)\right)\, ,
\end{equation}
where $\Sigma_{\rm d0}$ is the central face-on surface density of the
disk. $h$ is the radial exponential scale length, $x$  is an abbreviation for
$x=R/2h$, and $I$ and $K$ are Bessel functions (cf. Binney \& Tremaine 1987).
Finally, I consider a dark halo component, which I approximate as a
quasi-isothermal sphere,
\begin{equation}
\rho_{\rm h} = \rho_{\rm h0} \left(1 + \frac{r^2}{r_{\rm c,h}^2} \right)^{-1}
\end{equation}
with a rotation curve
\begin{equation}
v_{\rm c,h}^2(R) = 4\pi G \rho_{\rm h0} r_{\rm c,h}^2\left(1 -
\frac{r_{\rm c,h}}{R} arctan \frac{R}{r_{\rm c,h}}\right)\, .
\end{equation}

The free parameters to be determined by a fit of equation (3) to the observed
rotation curve are the central densities of bulge, disk, and dark halo,
respectively,
and the core radius of the dark halo. As is well known (see van Albada 1997 for
a comprehensive discussion), the decomposition of rotation curves is by no
means unique, but allows for a large variation of the parameters. The bulge
parameters, however, are well constrained, because they rely mainly on the
inner most data point in Figs.~2 to 5, where the disk and dark halo
contributions
are negligible. The central density of the bulge is determined as $\rho_{\rm
b0} = 5 M_\odot pc^{-3}$ which implies a total mass of the bulge of
$8 \cdot 10^{10} M_\odot$. First, I consider the `maximum disk' case, where
the disk contribution to the rotation curve is maximised. As can be seen from
Fig.~2 only a very minor dark halo, if any at all, is required to fit the
observed rotation curve. Next, I illustrate a `medium disk' decomposition
which requires a dark halo in Fig.~3. De Jong (1995) has shown that the
radial exponential scale lengths of galactic disks are usually smaller in the
infrared than at optical wavelengths. Since the infrared scale may be more
appropriate to describe the distribution of mass in the disk, I consider also
cases where the scale length of NGC\,488 has been reduced following de Jong
(1995) by a factor of 2/3. In Figs.~4 and 5 the corresponding `maximum disk'
and a `medium disk' decomposition of the rotation curve are shown. The
parameters derived from the various fits are summarized in Table 1. As can be
seen from Table 1 the dark matter contributes in most cases considerably
to the mass budget.

%
%
\begin{table}[t]
\caption{Disk and dark halo parameters}
\label{Parms}
\begin{center}
\begin{tabular}{cccccccc}
\hline
 & h & & $\Sigma_{\rm d0}$ & ${M_{\rm d}}^{\rm a}$ & $r_{\rm c,d}$ &
 $\rho_{\rm h0}$ & ${M_{\rm dh}}^{\rm a}$ \\[0.5ex] \hline
1 & 5.8 & max disk & 1900 & 21 & -- & 0     &  0\\
2 & 5.8 & med disk &  950 & 10 &  4 & 0.13  & 14\\
3 & 3.9 & max disk & 1450 & 10 & 10 & 0.055 & 15\\
4 & 3.9 & med disk &  800 &  6 &  5 & 0.14  & 20\\[0.5ex]
  & $kpc$ & & $M_\odot$ & $10^{10}$ & $kpc$ & $M_\odot$ & $10^{10}$\\
  &  & & $pc^{-2}$ & $M_\odot$ & & $pc^{-3}$ & $M_\odot$\\ \hline
\end{tabular}
\end{center}
\begin{list}{}{}
\item[$^{\rm a}$] Total mass within a radius of $10 kpc$.
\end{list}
\end{table}

\section{Discussion and Conclusions}

In order to discuss the viability of the dynamical disk models derived in the
previous section I consider two diagnostics.

First, I estimate the expected {\it vertical} scale height of the disk, even
though finite scale heights have been neglected in the decompositions of the
rotation curve. An isothermal self-gravitating stellar sheet follows a vertical
$sech^2(z/z_0)$ density profile (cf. Binney \& Tremaine 1987), with the
vertical scale height given by
\begin{equation}
z_0 = \frac{\sigma^2_{\rm W}}{\pi G \Sigma_{\rm d}}\, ,
\end{equation}
where $\sigma_{\rm W}$ is the dispersion of the vertical velocity components of
the stars and $\Sigma_{\rm d}$ denotes again the surface density of the disk.
Since the bulge is so massive its gravitational force field has to be taken
into account. In regions $R^2>>r_{\rm c,b}^2$ and $z^2<<R^2$, which are of
interest here, the vertical gravitational force field due to the bulge is
approximately given by
\begin{equation}
K_{\rm z,b}(R) = 22.0 \cdot G \rho_{\rm b0} \left(1 + \frac{R^2}{r_{\rm
c,b}^2}\right)^{-\frac{3}{2}} z\, .
\end{equation}
Fuchs \& Thielheim (1979) have considered the case of an isothermal sheet
imbedded in a linear force field. They show that the density profile of such a
disk follows still approximately a $sech^2(z/z_0\arcmin)$ law with a
modified scale height which can be expressed using equation (9) as
\begin{equation}
z_0\arcmin(R) = \frac{z_0(R) \Sigma_{\rm d}(R)}{\Big(
\Sigma_{\rm d}(R)  + 1.6 \rho_{\rm b0}
z_0\arcmin(R) \Big(1 + \frac{R^2}{r_{\rm
                    c,b}^2}\Big)^{-\frac{3}{2}}\Big)} \, .
\end{equation}
This can be solved explicetely for $z_0\arcmin(R)$.

Next, I consider the Toomre stability parameter $Q$ (cf. Binney \& Tremaine
1987),
\begin{equation}
Q = \frac{\kappa\sigma_{\rm U}}{3.36 G\Sigma_{\rm d}}\, ,
\end{equation}
where $\kappa$ denotes the epicyclic frequency, $\kappa^2 =
2(\frac{v_{\rm c}}{R})^2(1 + \frac{R}{v_{\rm c}} \frac{dv_{\rm c}}{dR})$, and
$\sigma_{\rm U}$
is the radial velocity dispersion of the stars. The velocity dispersions have
been directly observed in NGC\,488 by GKM and in equations (8) and (10) I use
the analytical fitting formulae given by the authors.
   \begin{figure}[htbp]
   \begin{center}
   \leavevmode
      \epsffile{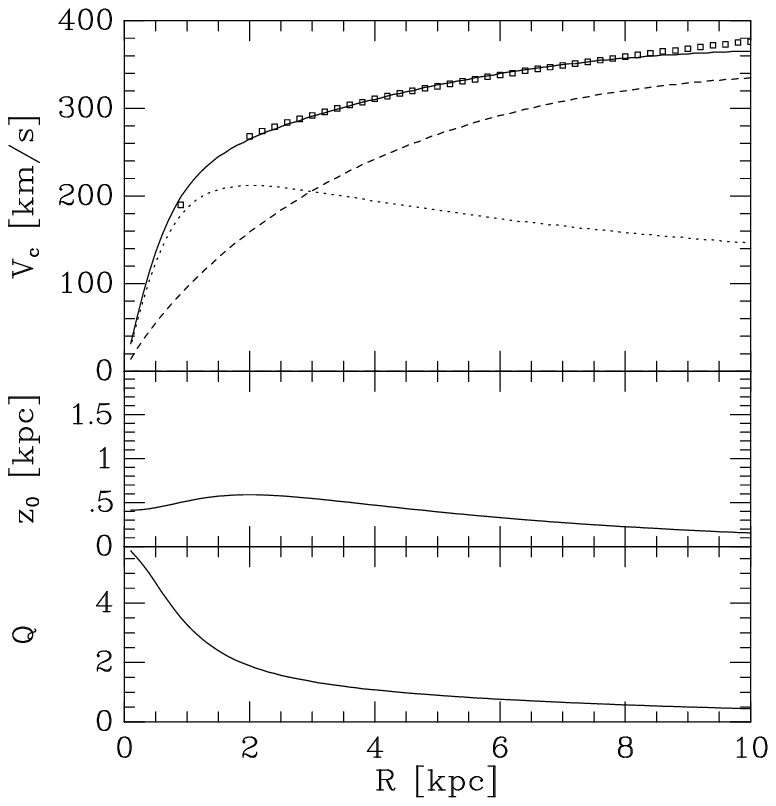}
\caption{Upper panel: Maximum disk decomposition of the rotation curve of
NGC\,488. The parametrization of the observed rotation curve by GKM is
shown by square symbols. The innermost data point comes from Peterson (1980).
The contributions by the bulge (dotted line) and the disk (short dashed line)
are indicated. Middle panel: Estimated vertical scale height of the disk.
Lower panel: Toomre stability parameter. }
         \label{disk1}
   \end{center}
   \end{figure}
\begin{figure}[htbp]
\begin{center}
   \leavevmode
      \epsffile{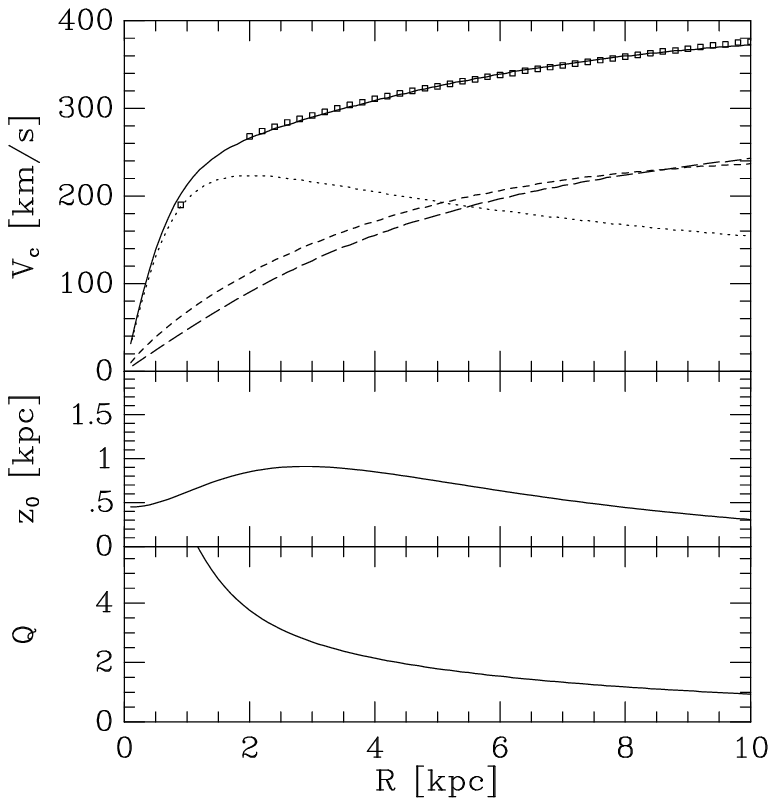}
\caption{Medium disk decomposition of the rotation curve of NGC\,488. The
contribution by the dark halo is shown as a long dashed line.}
\label{disk2}
\end{center}
\end{figure}
\begin{figure}[htbp]
\begin{center}
   \leavevmode
      \epsffile{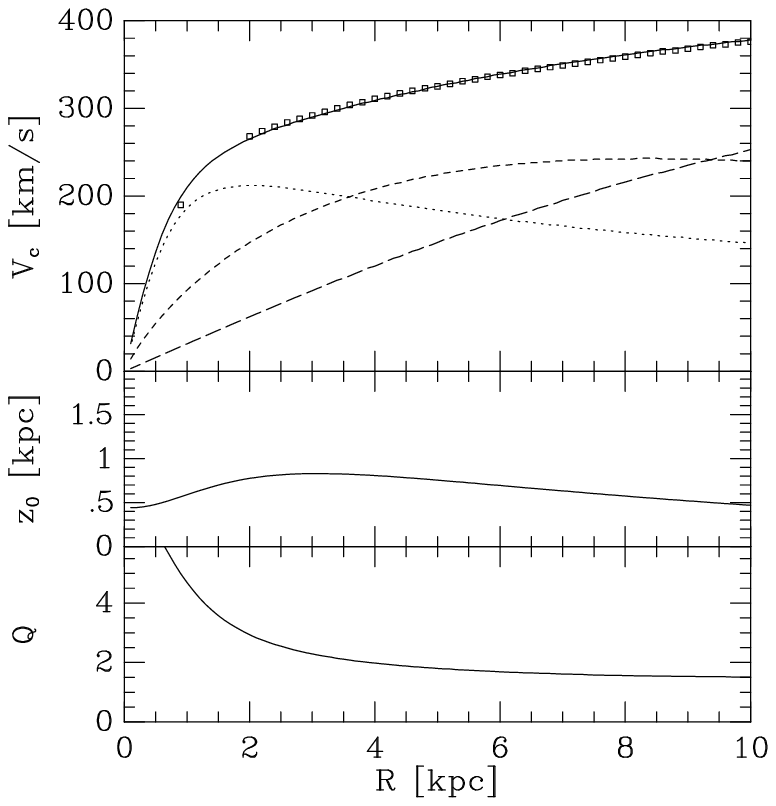}
\caption{Maximum disk decomposition of the rotation curve of NGC\,488 assuming
a radial scale length of the disk of 2/3 of the optical radial scale length.}
\label{disk3}
\end{center}
\end{figure}
\begin{figure}[htbp]
\begin{center}
   \leavevmode
      \epsffile{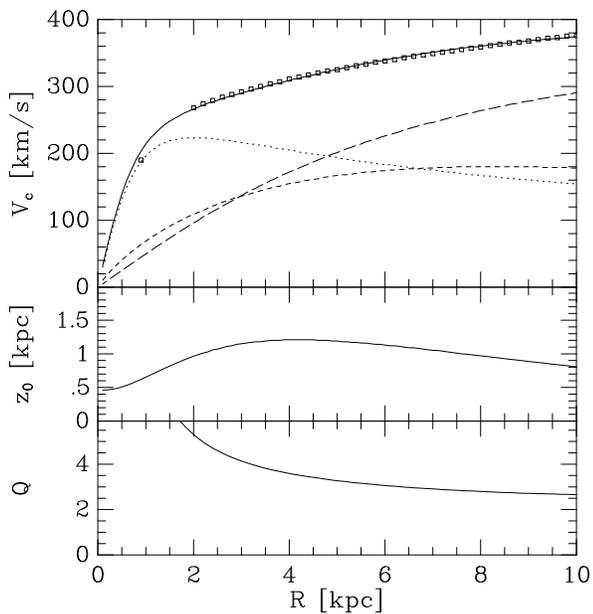}
\caption{Medium disk decomposition of the rotation curve of NGC\,488 assuming a
radial scale length of the disk of 2/3 of the optical radial scale length.}
\label{disk4}
\end{center}
\end{figure}

In the maximum disk case the derived vertical scale height is rather small,
$z_0/h$ = 0.07 at $R$ = 5 kpc. The ratio of vertical-to-radial scale
lengths is related to the flattening parameter of the disk by $q_0 =
0.7 \cdot z_0/h$ (Fuchs et al. 1996). Guthrie (1992) finds for the disks of Sa
galaxies typical values of $q_0 = 0.18 \pm 0.03$, implying $z_0/h = 0.25
\pm 0.04$. An even more severe argument against this disk model is a $Q$
parameter less than one. Such a disk will undergo violent gravitational
instabilities (Hockney \& Hohl 1969), whereas the very smooth optical
appearance of the disk (Elmegreen 1981) gives no indications for such
processes. The medium disk model is much more acceptable in this respect. Its
vertical scale height is about twice that of the maximum disk, but still
smaller than the values found by Guthrie (1992). The maximum disk model
assuming
a shorter radial scale length is very similar to the medium disk model with the
optical radial scale length; $z_0/h$ is 0.21 in this case. In the medium disk
model with the shorter radial scale length the ratio of vertical-to-radial
scale lengths is $z_0/h  =$ 0.31, which would be consistent with the upper
value given by Guthrie (1992).
However, the stability parameter is $Q \approx$ 3 which would
make the disk dynamically too hot to develop spiral structure, as can be seen,
for instance, in the numerical disk simulations by Sellwood \& Carlberg (1984),
whereas the optical image of NGC\,488 (Elmegreen 1981) shows clear regular
spiral arms. Thus I conclude that only models of the second or third type give
a reasonable description of the dynamics of the disk of NGC\,488.

This discussion shows that the disk of NGC\,488 must be dynamically cool.
Obviously this fact sets severe constraints on the merging history of NGC\,488.
The details of energy transfer from a satellite galaxy merging with its parent
galaxy to the disk of the parent galaxy are fairly complicated. They have been
studied
by T\'{o}th \& Ostriker (1992) and, in particular, by Quinn \& Goodman (1986),
Quinn, Hernquist \& Fullagar (1993), Walker, Mihos \& Hernquist (1996), and
Huang \& Carlberg (1997) in extensive numerical simulations. These simulations
have shown that a satellite galaxy looses its orbital energy due to a number of
effects. It induces coherent distortions of the disk of the parent galaxy such
as tilts, warps, and non--axisymmetric structures like bars and spiral arms,
which are not very effective in heating the disk dynamically. Some of the
energy of the satellite is carried away by its debris. Quinn, Hearnquist \&
Fullagar (1993) estimate that only about one half of the energy of the
satellite is converted to random kinetic energy of the disk of the parent
galaxy. Huang \& Carlberg (1997) have shown that the energy
transfer depends also on the structure of the satellite. Low density
satellites, for instance, do not reach the inner parts of the disk before they
are
disrupted. They heat only the outer part of the disk or the inner halo, but
do not contribute to the bulge. Observational evidence for such merging events
in nearby spiral galaxies is discussed by Zaritzky (1995). GKM have observed
the kinematics of the inner part of the disk of NGC\,488.
The numerical simulations show that satellites which reach these parts of
the disk will be on orbits which are lowly inclined with respect to the
plane of the parent galaxy. The energy which the satellites loose while
they spiral inwards will be then deposited mainly into the disk until they have
reached the bulge, because the disk density is dominating as can be shown
using the
parameters given in Table 1. In order to illustrate the disk heating effect I
consider the strip of the disk ranging from galactocentric radii 5 to 10 kpc.
It is straightforward to calculate from the centrifugal forces modelled in
equations (4) to (7) the loss of potential energy $\Delta E$ of a satellite
spiralling from radius 10 kpc to 5 kpc. Using the parameters given in Table 1 I
find for all disk models about the same value of $\Delta E = 8.4 \cdot 10^4
(km/s)^2 \cdot M_{\rm s}$, where $M_{\rm s}$ denotes the mass of the satellite.
This energy
loss will lead to an increase of the peculiar velocities of the disk stars,
which is approximately given by (Ostriker 1990),
\begin{equation}
\frac{1}{2}
\Delta E \approx M_{\rm d} \delta v^2 \approx \frac{1}{3} M_{\rm d} \delta
\sigma_{\rm U}^2\, ,
\end{equation}
where $M_{\rm d}$ is the disk mass contained in the strip from 5 to 10 kpc
radius. Equation (12) may be cast into the form
\begin{equation}
\frac{M_{\rm s}}{M_{\rm d}} = \frac{2}{3}\frac{\sigma_{\rm U}^2}{(\frac{\Delta
E}{M_{\rm s}})}
\frac{\delta Q^2}{Q^2} \, .
\end{equation}
Averaging the observed velocity dispersion $\sigma_U$ radially and assuming a
pre-merger stability parameter of $Q = 1$, I obtain for disk models 2 and 3
mass estimates of $M_{\rm s} = 8 \cdot 10^8 M_\odot$,
meaning that the present dynamical
state of the disk of NGC\,488 is consistent with {\it one} merging event in the
past with a satellite galaxy of such a mass. The build-up of the entire bulge
of NGC\,488, on the other hand, would have required about 100 of such mergers.

Reshetnikov \& Combes (1997) have argued that stellar disks will be cooled
dynamically after a merging event by newly born stars on low - velocity -
dispersion orbits. Fresh interstellar gas maintaining a high star formation
rate is supposed to be replenished in the inner parts of the disks by radial
gas inflow which is induced by non-axisymmetric tidal
perturbations during the merging event. The gas consumption rate required to
cool the disk can be estimated in the following way. If, for instance, the
surface density of the disk has grown after a certain time interval to
$(1+\epsilon)\Sigma_{\rm d}$, the stability parameter is roughly given by
\begin{equation}
Q^2 = \frac{\kappa^2(\sigma_{\rm U}^2\Sigma_{\rm d} + \sigma_{\rm
U,i}^2\epsilon\Sigma_{\rm
d})}{(3.36 G (1+\epsilon) \Sigma_{\rm d})^2 (1+\epsilon) \Sigma_{\rm d}\, ,}
\end{equation}
where I approximate the velocity dispersion of the disk by a mass weighted
average of the velocity dispersion of the heated stellar component and the
velocity dispersion $\sigma_{\rm U,i}$ of the cooling stars. The stability
parameter (14) can be expressed in terms of the stability parameter of the
heated disk, $Q_{\rm h}$, as
\begin{equation}
Q^2 = Q_{\rm h}^2 \frac{1+\epsilon(\frac{\sigma_{\rm U,i}}{\sigma_{\rm
U}})^2}{(1+\epsilon)^3} \approx Q_{\rm h}^2(1+\epsilon)^{-3}\, ,
\end{equation}
where I have neglected the velocity dispersion of the newly born stars. Thus,
if a cooling of the stability parameter by an amount of $\delta Q^2$ is
necessary, this requires a relative increase of the disk mass of
\begin{equation}
\epsilon = \left( 1-\frac{\delta Q^2}{Q_{\rm h}^2} \right)^{-1/3} - 1
\end{equation}
in the form of newly born stars. If, for instance,
$\epsilon Q^2/Q_{\rm h}^2 = 0.5$ this implies $\delta = 0.26$. This is of the
order of what is available in NGC\,488 in the form of interstellar gas.
The total mass of hydrogen in
atomic and molecular form in NGC\,488 is estimated as $2.5 \cdot 10^{10}
M_\odot$ (Young et al.~1996), which after multiplication with a factor of 1.4
in order to account for heavier elements has to be compared with the stellar
disk mass estimates given in Table 1. The gas content of dwarf spirals is
typically of the order of $10^9 M_\odot$ (Broeils 1992), i.e.~only about one
tenth of the mass needed to cool the disk after a merging event.
From this discussion I conclude
that the disk of a Sa galaxy may in principle `recover' from a minor merger and
flatten its disk again and develop again spiral structure, although this has
still to be studied in detail. The build-up of the
bulge of NGC\,488, however, would require about 100 mergers implying a merger
rate of 10 mergers per Gyr if a steady merger rate is assumed. This would mean
that
in NGC\,488 at present about five times the stellar disk mass must be converted
per Gyr from interstellar gas into stars in order to keep the stellar disk in
its present dynamical state. The actually observed star formatiom rate (Young
et al.~1996), on the other hand, is only $1.2 \cdot 10^9 M_\odot Gyr^{-1}$.
So it seems unlikely that the bulge of NGC\,488 has been build up entirely by
accretion of satellite galaxies.

%
%
\begin{acknowledgements}
      I am grateful to A.~Broeils, R.~Wielen and the referee for helpful
discussions and suggestions.
\end{acknowledgements}

\end{document}